\documentclass[twocolumn,preprintnumbers,secnumarabic,amsmath,amssymb,superscriptaddress,nofootinbib]{revtex4}

\newcommand\T{\rule{0pt}{2.6ex}}
\newcommand\B{\rule[-1.2ex]{0pt}{0pt}}

\def\beq{\begin{equation}}
\def\eeq{\end{equation}}
\def\bea{\begin{eqnarray}}
\def\eea{\end{eqnarray}}
\def\d{{\rm d}}

\begin{document}


\title{Compactification Effects in D-brane Inflation}

\medskip\
\author{Daniel Baumann}

\author{Anatoly Dymarsky}
\affiliation{%
School of Natural Sciences,
 Institute for Advanced Study,
Princeton, NJ 08540}

\author{Shamit Kachru}
\altaffiliation{On leave of absence from Stanford University and SLAC.}
\affiliation{Kavli Institute for Theoretical Physics, Santa Barbara, CA 93106}

\author{Igor R.~Klebanov}
\affiliation{Department of Physics and PCTS, Princeton University,
Princeton, NJ 08544}

\author{Liam McAllister}%
\affiliation{%
Department of Physics, Cornell University,
Ithaca, NY 14853}

\date{\today}

\date{\today}
\begin{abstract}
\vskip .1cm
\begin{center}{{\emph{Dedicated to the memory of Lev Kofman}}}
\end{center}
\vskip .1cm
We determine the scalar potential for a D3-brane in a warped conifold background subject to general ultraviolet perturbations.   Incorporating the effects of imaginary anti-self-dual (IASD) fluxes and four-dimensional curvature at the nonlinear level, we compute the leading terms in the D3-brane potential.   We then provide strong cross-checks of our results by reproducing them in the dual gauge theory.  Finally, we observe that the D3-brane potential induced by nonperturbative effects on D7-branes can be represented by a ten-dimensional supergravity solution containing suitable IASD fluxes.  Our method allows for the systematic inclusion of compactification effects and serves to constrain the D3-brane effective action in a large class of stabilized compactifications.
\end{abstract}

\maketitle

{\bf Introduction.}\
The success of inflation~\cite{Inflation} depends sensitively on Planck-suppressed contributions to the inflaton action~\cite{Baumann:2008aq}.  This strongly motivates realizing inflation in an ultraviolet-complete theory, such as string theory, and then computing these contributions.

In this Letter we study compactification effects in D-brane inflation~\cite{DvaliTye,Burgess}.  A concrete arena for D-brane inflation is a finite warped throat attached to a compact space~\cite{KKLMMT}.  Computing the D3-brane action in the noncompact limit is straightforward, as explicit supergravity solutions such as \cite{KS} are available for noncompact throats.  However,  on general grounds one expects that objects and fields in the bulk 
of the compactification, {\it i.e.}~outside the warped throat,
can induce order-unity corrections to the inflationary slow-roll parameters.  D-branes, orientifold planes, fluxes, and quantum effects in the bulk may not preserve the same supersymmetry as a D3-brane, and hence can contribute to the D3-brane potential. These `compactification effects' must be incorporated, or shown to be suppressed, in any explicit realization of D3-brane inflation.

We propose a systematic approach to this problem.  Taking the compact bulk to be a rather general solution of type IIB supergravity, we determine how the form of this solution in the ultraviolet region of a finite throat affects the potential for a D3-brane well inside the throat.  Our strategy is to study a noncompact solution subject to general deformations, as a computable proxy for a finite throat attached to a compact space with moduli-stabilizing ingredients.

A primary goal of our work is a systematic characterization of the structure of the effective potential, as captured by the spectrum of  scaling dimensions $\Delta_i$ in
\beq
\label{equ:1}
V(\phi) = \sum_i c_i \, \frac{\phi^{\Delta_i}}{M_{\rm UV}^{\Delta_i - 4}}\, .
\eeq
Here $\phi$ is the canonically-normalized field related to the D3-brane position and $M_{\rm UV}$ is a scale determined by the location at which the throat is glued to the bulk.

Our analysis incorporates
imaginary anti-self-dual (IASD) three-form fluxes and four-dimensional curvature in a unified framework.
We provide general solutions for IASD fluxes on arbitrary Calabi-Yau cones and derive the flux-induced contributions to the D3-brane potential in closed form.
Specializing to the singular conifold, we determine the dimensions $\Delta_i$ in (\ref{equ:1}).
The inclusion of the nonlinear effects of fluxes and 
curvature is a substantial step beyond our linear analysis in~\cite{Baumann2009}.

A nontrivial check of our results follows from using the AdS/CFT correspondence~\cite{MAGOO} to represent non-normalizable flux perturbations
as perturbations to the Lagrangian of the dual
conformal field theory (CFT).  Building on the work of Ceresole {\it et al.}~\cite{Ceresole2000}, we provide the
operator in the conifold CFT dual to each mode of flux.  In the case of chiral operators perturbing the superpotential, we compute the potential in the CFT and find agreement with the supergravity result~\cite{LongPaper}.

Our approach is strongly reminiscent of a four-dimensional effective field theory analysis, but the field theory that governs the D3-brane potential is strongly coupled.  Furthermore, some of the contributing operators have irrational dimensions and are difficult to study on the field theory side.  Our method, which is to consider
general perturbations of the ultraviolet region of the supergravity solution, uses AdS/CFT to provide a tractable problem that realizes the spirit of the four-dimensional effective field theory approach.

We limit our presentation in this Letter to explicit results for chiral operators; extensions to non-chiral operators and further computational details will appear in a companion paper~\cite{LongPaper}.

\vskip 4pt
{\bf Equations of motion.}\
\label{sec:10dSUGRA}
For the line element, we use the ansatz
\beq
\label{equ:warping}
\d s^2 = e^{2A(y)} g_{\mu \nu} \d x^\mu d x^\nu + e^{- 2 A(y)} g_{mn} \d y^m \d y^n\, ,
\eeq with $g_{mn}$ a metric that is Calabi-Yau at leading order.
We take
the five-form flux to be
$\tilde F_5 =  (1+ \star_{10})\, \d \alpha(y) \wedge {\rm Vol}_4$.
D3-branes couple to a particular combination of fields,
\beq
V_{\rm D3} = T_3 \left( e^{4A} - \alpha \right) \equiv T_3 \Phi_- \, ,
\eeq
where $T_3$ is the D3-brane tension.
In the following we are therefore interested in perturbations of the scalar quantity $\Phi_- \equiv e^{4A} - \alpha$. 
Furthermore,
it will be convenient to define IASD three-form fluxes,
\beq
G_- \equiv (\star_6-i) G_3 \quad {\rm and} \quad  \Lambda \equiv e^{4A} G_{-}\, ,
\eeq
where $\star_6$ is the six-dimensional Hodge star. Expanding around backgrounds with $\Phi_-=G_{-}=0$, these fluxes obey~\cite{LongPaper}
\beq
\label{equ:fluxEoM}
\d \Lambda =0   \quad {\rm and} \quad  \star_6 \Lambda = - i \Lambda \, .
\eeq
Finally, the D3-brane potential is determined by
\beq
\label{equ:MasterX}
\nabla^2 \Phi_- = \frac{g_s}{96} | \Lambda|^2 + {\cal R}_4 \, ,
\eeq
where $\nabla^2$ is the Laplacian constructed from $g_{mn}$,
$g_s$ is the string coupling, and ${\cal R}_4$ is the four-dimensional Ricci scalar.
 
\vskip 4pt
{\bf IASD fluxes in Calabi-Yau cones.}\
In \cite{Baumann2009} we studied the homogeneous solutions of (\ref{equ:MasterX}), ignoring the effects of IASD fluxes and curvature.
The solutions to (\ref{equ:MasterX}) were then harmonic functions, $\Phi_- = f$, which satisfy $\nabla^2 f=0$.
In this Letter we extend this to a nonlinear treatment, emphasizing that IASD fluxes can make important contributions to the D3-brane potential.

We first present solutions to the flux equations of motion in a general Calabi-Yau cone.
To construct solutions to (\ref{equ:fluxEoM}), we use the K\"ahler form $J$, the holomorphic $(3,0)$ form $\Omega$, and harmonic functions $f$
as building blocks.
Three distinct types of closed, IASD three-forms can be constructed using the above ingredients.  In \cite{LongPaper} we give the most general such solutions, while here we specialize to solutions
dual to perturbations by chiral operators,
in which case the harmonic functions $f$ are in fact holomorphic and the fluxes are of pure Hodge type.

The  first flux series is of Hodge type $(1,2)$,
\beq
\label{equ:LambdaI}
\Lambda_{\rm I} = \nabla \nabla f_1 \cdot \bar \Omega\, ,
\eeq
which in components is $(\Lambda_{\rm I})_{\alpha\bar\beta\bar\gamma}=\nabla_\alpha \nabla_\sigma f_1 \, g^{\sigma \bar \zeta} \, \bar \Omega_{\bar \zeta \bar \beta \bar \gamma}$.
The second flux series is non-primitive $(2,1)_{\rm NP}$,
\beq
\label{equ:LambdaII}
\Lambda_{\rm II} = \partial f_2 \wedge J\, .
\eeq
The third flux series is of type $(3,0)$,
\beq
\label{equ:LambdaIII}
\Lambda_{\rm III} =f_3 \, \Omega \, .
\eeq

{\bf Spectrum of the D3-brane potential.}
\label{sec:spectrum}
Next, we derive the spectrum of contributions to the D3-brane potential.
We first study the flux source terms in~(\ref{equ:MasterX}), then incorporate
${\cal{R}}_{4}$ as a source.
From (\ref{equ:MasterX}) -- (\ref{equ:LambdaIII}) we obtain the flux-induced potential
\beq \label{phiminus}
\Phi_-  = \frac{g_s}{32} \left[ g^{\alpha \bar \beta} \nabla_\alpha {f}_1 \, \overline{\nabla_\beta {f}_1} + 2 |{f}_2|^2 +  2\, \nabla^{-2} |{f}_3|^2 \right]\, ,
\eeq
up to the addition of a harmonic function.
The result (\ref{phiminus}) is completely general, and is valid for IASD fluxes on an arbitrary Calabi-Yau cone.
For concreteness we now specialize to
the singular conifold~\cite{KW}.

The harmonic functions on the conifold are defined by their quantum numbers $ \{ j_1, j_2, R_f \}$ under the $SU(2)\times SU(2)\times U(1)_R$ isometries of $T^{1,1}$~\cite{Ceresole2000}.
The radial scaling dimensions $\Delta_f$ are 
\bea
\Delta_f &\equiv& - 2 + \sqrt{H(j_1, j_2, R_f)+4}  \\
&=& \frac{3}{2}\ , \ 2 \ , \ 3 \ , \ \sqrt{28} -2 \ , \ \cdots \label{equ:homogeneous}\, ,
\eea
where $H(j_1,j_2,R_f) \equiv 6 \left[j_1(j_1+1) + j_2 (j_2+1) - R_f^2/8  \right]$.
For chiral modes, $j_1=j_2=\frac{1}{2} R_f$, we have $\Delta_f = \frac{3}{2} R_f$.

With this information about the spectrum of harmonic functions on $T^{1,1}$, the scaling dimensions and charges of the three series of flux modes can be determined \cite{LongPaper}.
The contributions to the potential 
are of the form  (\ref{equ:1}) with
$\Delta_{ij} \equiv \Delta_i + \Delta_j -4$,
where $\Delta_i$ and $\Delta_j$ are the scaling dimensions of the individual fluxes~\cite{LongPaper}. The result is
\beq
\label{equ:DeltaL}
\Delta_\Lambda \ =\  1\ , \ 2\ ,\ \frac{5}{2}\ , \ \sqrt{28} - \frac{5}{2} \  , \ \cdots
\eeq

The final term shown, with dimension $\sqrt{28}-\frac{5}{2} \approx 2.79$, is the leading contribution of an operator in a long multiplet.  This term is not protected by supersymmetry or by a global symmetry, and hence would be difficult to obtain in a field theory analysis at large 't~Hooft coupling,
or in a four-dimensional supergravity analysis like \cite{BDKMMM,BDKM} in which only superpotential interactions are computed.
We expect that this mode of flux will contribute to the D3-brane potential in a general compactification, and may in some cases play an important role in the inflationary phenomenology. 
A substantial advantage of our supergravity approach is the ability to capture low-dimension non-chiral contributions of this sort.

A D3-brane in a compact space receives additional contributions to its potential when the four-dimensional Ricci scalar is nonvanishing: one must then solve (\ref{equ:MasterX})
with the Ricci scalar given by
${\cal R}_4  \approx \frac{4}{M_{\rm pl}^2} \bigl(V_0 + T_3\Phi_-\bigr)$.
Here we have assumed a quasi-de Sitter phase sourced
by the D3-brane potential $T_3\Phi_-$ and by the effects of other sectors included in the constant $V_{0}$.
We observe that the source term in (\ref{equ:MasterX}) proportional to $V_{0}$
gives rise to precisely the `eta problem' mass term in \cite{KKLMMT} ({\it cf.}~\cite{Buchel}).
The complete series of corrections induced by the Ricci scalar is~\cite{LongPaper}
\beq
\label{equ:DeltaR}
\Delta_{\cal R} \ =\  2 \  , \ 3\ , \ \frac{7}{2} \ , \ \cdots
\eeq
Although for brevity we have focused here on the radial scaling dimensions $\Delta$, one can use the above methods to determine the angular dependence of each term \cite{LongPaper}.

\vskip 4pt
{\bf CFT duals of flux perturbations.}\
A useful dual formulation of our results is in terms of perturbations to the Lagrangian of the dual CFT~\cite{KW}.  This is an ${\cal N}=1$ superconformal theory
with gauge group ${\cal G} = SU(N) \times SU(N)$ and continuous global symmetries $G = SU(2) \times SU(2) \times U(1)_{R}$.
The matter content consists of chiral superfields $A_i$ and $B_j$ ($i,j=1,2$) which are in the ($N$, $\bar N$) and ($\bar N$, $N$) of ${\cal G}$ and in the $(\frac{1}{2}, 0, \frac{1}{2})$ and $(0,\frac{1}{2},\frac{1}{2})$ of $G$.
The chiral gauge field strength superfields $W_\alpha^{(1,2)}$ have dimension $\Delta = \frac{3}{2}$ and $R$-charge $R=1$, while the $A$ and $B$ fields have $\Delta = \frac{3}{4}$ and $R=\frac{1}{2}$.

Our goal is to determine the most relevant operators in the CFT that contribute to the potential on the Coulomb branch. We focus primarily on chiral operators, deferring a more general treatment to \cite{LongPaper}.
For the CFT in question there are three series of chiral operators \cite{Ceresole2000}: 
\bea
{\cal O}_{\rm I} &=& {\rm Tr}(AB)^k\ , \label{equ:O1}\\
 {\cal O}_{\rm II}^\alpha &=& {\rm Tr}[(W^\alpha_{(1)} + W^\alpha_{(2)}) (AB)^k]\ , \\
  {\cal O}_{\rm III} &=& {\rm Tr}[(W_{(1)}^2+ W_{(2)}^2)(AB)^k]\, , \label{equ:O3}
\eea
where the integer $k$ obeys $k\ge 1$ for series I and II and $k\ge0$ for series III.

We are interested in compactification effects that break the supersymmetry of the CFT.  This is efficiently represented in terms of spurions $X$ and ${X}^{\alpha}$, {\it i.e.}~non-dynamical
fields whose expectation values break supersymmetry, $\int  \hspace{-0.05cm}d\theta_{\alpha}{X}^{\alpha}  \neq 0$, $\int  \hspace{-0.05cm}d^2 \theta X \neq 0$, with the remaining components vanishing.
The operators dual to IASD flux perturbations can then be written as 
\beq
\Delta {\cal L} \, =\, \int d^2 \theta \, \left[ {\cal O}_{\rm I} \,+ \,  {\cal O}_{\rm II}^\alpha {X}_{\alpha}  \,+ \,  {\cal O}_{\rm III} X\right] \ \ + \ \ c.c. \ ,
\eeq which selects the top  ($\theta^2$) component of ${\cal O}_{\rm I}$, the middle component of ${\cal O}_{\rm II}^{\alpha}$, and the bottom component of ${\cal O}_{\rm III}$.
(In \cite{Baumann2009} we studied linearized perturbations $\int  \hspace{-0.05cm}d^4 \theta \, {\cal O} X^{\dagger}X$; these are dual to harmonic modes of $\Phi_-$, whose spectrum is given in (\ref{equ:homogeneous}).)

We therefore obtain the following matching between the components of the supermultiplets of operators (\ref{equ:O1}) -- (\ref{equ:O3}) and the IASD flux perturbations (\ref{equ:LambdaI}) -- (\ref{equ:LambdaIII}):
\begin{table}[h!b!p!]
\begin{tabular}{l l |c |c c c}
\hline
\hspace{0.15cm} {\bf Flux} \T \B   &\hspace{.7cm}  {\bf Operator} & \hskip 4pt $\mathbf{\Delta}$ \hskip 4pt & $R$ & $j_1$ & $j_2$\hskip 2pt \\
\hline
\hskip 2pt $\nabla \nabla f_1 \cdot \bar \Omega$ \T  & \hskip 2pt $\int  \hspace{-0.05cm}d^2 \theta\, {\rm Tr}(AB)^k$ & \hskip 2pt \footnotesize{$\frac{3}{2} k + 1$} \hskip 2pt & \hskip 2pt \footnotesize{$k-2$} \hskip 2pt & \footnotesize{$\frac{1}{2}k$} & \footnotesize{$\frac{1}{2} k$} \hskip 2pt\\
\hskip 2pt  $\partial f_2 \wedge J$ \T \B &  $  \hskip 2pt  \int  \hspace{-0.05cm}d \theta_\alpha\, {\rm Tr}[W^\alpha_+ (AB)^k]$ \hskip 4pt&
\footnotesize{$\frac{3}{2} k + 2$}& \footnotesize{$k$} & \footnotesize{$\frac{1}{2} k$} & \footnotesize{$\frac{1}{2} k$} \hskip 2pt\\
\hskip 2pt $f_3  \Omega$\B &  \hskip 2pt  ${\rm Tr}[(W^2_+ (AB)^k]$  \hskip 4pt  & \footnotesize{$\frac{3}{2} k + 3$} & \footnotesize{$k+2$} & \footnotesize{$\frac{1}{2} k$} & \footnotesize{$\frac{1}{2} k$} \hskip 2pt\\
\hline
\end{tabular}
\end{table}

$\bullet$
The perturbations $\int  \hspace{-0.05cm}d^2 \theta \, {\cal O}_{\rm I}$
yield superpotential perturbations of the CFT and are dual to perturbations by $G_{(1,2)}$ fluxes 
in series~I.  This is consistent with the results of \cite{PolchinskiStrassler, CIU}, where it was found that $G_{(1,2)}$ flux generates superpotential interactions for a D3-brane.
We remark that the leading $(\Delta=1)$ term in (\ref{equ:DeltaL}) corresponds to the lowest-dimension superpotential perturbation $\int \hspace{-0.05cm}d^2 \theta \, {\rm Tr}(AB)$ and gives the dominant contribution in~\cite{BDKM} (see (\ref{equ:Fterm}) below).

$\bullet$
The operators $\int  \hspace{-0.05cm}d \theta_\alpha\,  {\cal O}_{\rm II}^\alpha$ correspond to the non-primitive $G_{(2,1)}$ fluxes 
of series~II.
This agrees with the results of \cite{PolchinskiStrassler, CIU}, where it was shown that non-primitive $G_{(2,1)}$ flux couples the gaugino to the matter fermions. 

$\bullet$
Finally, the operators ${\cal O}_{\rm III}$ are mapped to $G_{(3,0)}$ fluxes 
in series III. This is consistent with the fact that $(3,0)$ flux creates a mass term for the gaugino \cite{PolchinskiStrassler}.

\vskip 4pt
For perturbations in series I and II, the leading terms in the CFT scalar potential are governed by holomorphic data and by the unperturbed K\"ahler potential, and one can show exact agreement between the potential on the CFT side and in supergravity \cite{LongPaper}. For operators in series III, K\"ahler potential perturbations render the potential uncomputable in the CFT, but one can still exhibit strong consistency checks \cite{LongPaper} of the correspondence to series III fluxes.

Note that perturbing by a relevant operator, $\Delta {\cal L} \, =\, c\, {\cal{O}}$, does not create an instability in the infrared if $c$ is sufficiently small in units of the hierarchy of scales in the throat.  Requiring that the D3-brane energy does not drive decompactification leads to a similar constraint on~$c$~\cite{Baumann2009,LongPaper}.  We consistently study the $c \ll 1$ regime in which instabilities are absent.

\vskip 4pt
{\bf Superpotentials from fluxes.} \
\label{sec:4dSUGRA}
In the presence of a nonperturbative superpotential from gaugino condensation on wrapped D7-branes, a D3-brane feels a nontrivial potential \cite{BDKMMM}. We now point out that this potential can be represented in ten dimensions by IASD fluxes.  Consider a stack of $n$ D7-branes on a four-cycle 
defined by a holomorphic embedding condition $h(z_a) = 0$, where $z_a$ are complex coordinates on the conifold.
The superpotential of the D7-brane gauge theory depends on the D3-brane position via \cite{BDKMMM}
\beq
W_{\rm np}(z_a) = {\cal A}_0 h(z_a)^{1/n}\,e^{- a \rho} \equiv  {\cal A}(z_a) e^{- a \rho}\, ,
\eeq
where $\rho$ is the K\"ahler modulus associated with the volume of the 4-cycle and $a \equiv \frac{2\pi}{n}$.

Combining $W_{\rm np}(z_a)$ with the tree-level superpotential and K\"ahler potential, $K = - 3 \ln[\rho + \bar \rho - \gamma k(z_a, \bar z_a)] $, and working to leading order  in an expansion in $\frac{\gamma k}{\sigma} = \frac{\phi^2}{3M_{\rm pl}^2}$, with $2 \sigma \equiv \rho + \bar \rho$, one finds the following D3-brane potential~\cite{BDKM}:
\bea
\label{equ:Fterm}
V &=& \frac{\kappa^2}{12 \sigma^2} \frac{e^{-2a\sigma}}{\gamma} \Bigl( g^{\alpha \bar \beta} {\cal A}_\alpha \bar {\cal A}_{\bar \beta} + 2a \gamma (a \sigma  + 3) {\cal A} \bar {\cal A} \nonumber \\
&& - a \gamma(\bar {\cal A} g^{\alpha \bar \beta} k_{\bar \beta} {\cal A}_\alpha + c.c.) \Bigr) \ + \ {\rm harmonic}\, .
\eea
Comparing (\ref{phiminus}) to (\ref{equ:Fterm}) suggests the matching conditions
\beq
f_1 = c_1 {\cal A}\, ,  \qquad
f_2 = c_2 {\cal A} + c_3 k^\beta {\cal A}_\beta\, , \qquad
f_3 = 0\, .
\eeq
Choosing $c_1 =  c$, $c_2 = c\, \sqrt{a \gamma (a \sigma + 3)} $, and 
$2 c_2 c_3 = - c^2 a \gamma$,
with $c^2 \equiv \frac{\kappa^2}{\sigma^2} \frac{e^{-2a\sigma}}{\gamma T_3} \frac{8}{3g_s}$,
we recover (\ref{equ:Fterm}), at leading order in expansions in $\frac{\gamma k}{\sigma}$ and $\frac{1}{a\sigma}$.

\vskip 4pt
{\bf Towards a D7-brane geometric transition.}\
\label{sec:speculations}
It is tempting to speculate about the implications of the above result for a ten-dimensional representation of the nonperturbative superpotential associated with D7-branes ({\it cf.}~\cite{Frey:2005zz, Koerber:2007xk}).

Suppose that $n$ D7-branes  wrap a rigid four-cycle that is small compared to the rest of the compact space, {\it e.g.}~a small del Pezzo surface. Far from the D7-branes, it should be possible to
encode the effects of gaugino condensation in a local solution of ten-dimensional supergravity containing appropriate IASD fluxes.
It is therefore natural to conjecture that  nonperturbative effects on D7-branes actually {{\it source}} these fluxes.
This is somewhat similar to the geometric transition of D5-branes wrapping rigid curves, in which the resulting gaugino condensate is captured by three-form fluxes in a deformed geometry~\cite{KS, Vafa, MN}.

Specifically, we propose \cite{LongPaper} that a coupling between the D7-brane gauginos and the closed string background serves to source IASD fluxes when $\langle\lambda\lambda\rangle \neq 0$.  
As shown in detail in \cite{LongPaper}, the gaugino mass term of \cite{CIU} contributes a local source term, proportional to $\langle\lambda\lambda\rangle$, to the flux equation of motion. The resulting solution contains $G_{(1,2)}$ flux that precisely encodes the gaugino condensate superpotential.

Finally, we remark on an interesting implication of our result.
In \cite{Harvard} it was shown that the noncommutative superpotential induced on coincident D3-branes by three-form fluxes solves a problem in F-theory model-building by increasing the rank of the Yukawa matrix.  However, in \cite{Martucci} it was argued that a noncommutative superpotential should {\it{not}} arise at tree level, but should instead be induced by nonperturbative effects on D7-branes.  Our proposal yields an explicit connection between these works by representing the nonperturbative superpotential of \cite{Martucci} as the flux of \cite{Harvard}.

\vskip 4pt 
{\bf Conclusions.}\
We have presented a systematic method for obtaining the structure of the scalar potential of a D3-brane in a noncompact Calabi-Yau cone subject to arbitrary ultraviolet deformations, in an expansion around a background with $G_{-} = \Phi_- = 0$.  For the case of the conifold, we obtained the explicit spectrum of scaling dimensions for the leading correction terms, (\ref{equ:homogeneous}) -- (\ref{equ:DeltaR}), incorporating harmonic perturbations of $\Phi_-$, IASD fluxes, and four-dimensional curvature.  This result includes all significant contributions to the D3-brane potential discussed in the literature so far, and does so in a single coherent framework, classical ten-dimensional supergravity.

Our results constitute significant progress towards determining the Planck-suppressed contributions to the effective action in a large class of string compactifications.  By decoupling gravity in a noncompact configuration, then systematically reincorporating compactification effects, we have provided an approach in which the structure of the Planck-suppressed contributions can be computed.

\vskip 4pt
{\bf Acknowledgements.}\
We thank J.~Gray, Z.~Komargodski, J.~Maldacena, E.~Pajer, N.~Seiberg, and G.~Torroba for discussions.
This work was supported in part by the NSF under grants PHY-0855425 (D.B.), PHY-05-51164 (S.K.), PHY-0756966 (I.R.K.), and PHY-0757868 (L.M.),
by the DOE under grants DE-FG02-90ER40542 (A.D.) and DE-AC02-76SF00515 (S.K.), and by the Stanford Institute for Theoretical Physics (A.D. and S.K.).

\vfil

\begingroup\raggedright
\endgroup

\end{document}